\documentclass[amsmath,aps,prl,twocolumn]{revtex4}%,twocolumn
\usepackage{amsthm,amsfonts,graphicx,verbatim}

\newcommand{\Tr}{{\rm Tr}}

\newcommand{\D}{{\rm d}}

\newcommand{\be}{\begin{equation}}
\newcommand{\ee}{\end{equation}}
\newcommand{\bea}{\begin{eqnarray}}
\newcommand{\eea}{\end{eqnarray}}

\newcommand{\la}{\langle}
\newcommand{\ra}{\rangle}

\renewcommand{\epsilon}{\varepsilon}
\def\nn{\nonumber\\}

\usepackage{color}
\definecolor{Red}{rgb}{0.8,0,0}
\definecolor{Black}{rgb}{0,0,0}

\begin{document}
\title{The Entanglement of a Quantum Field with a Dispersive Medium}
\author{Israel Klich}
\affiliation{Department of Physics, University of Virginia, Charlottesville, Virginia 22904, USA}
%\\ (2) Department of Physics.... }
\begin{abstract}
%The modification of the quantum state of electromagnetic fields in the presence of matter yields fascinating effects such as the Casimir effect and the Lamb shift. Here we consider another aspect of such interaction: the entanglement of the matter and radiation as measured by the von-Neumann entropy of field. %Indeed, in the presence of a dissipative medium, the electromagnetic field cannot be described as a pure state anymore.
%The quantization of the electromagnetic field in the presence of a dissipative medium results in a mixed state, and as such carry entropy. 
%results in modification  Casimir effect. 
%When the objects are dissipative, the Casimir free energy is modified. In a similar way, dissipation causes the field to be in a mixed state, and as such carry entropy. 
%Here, we study the contribution of dissipation and dispersion in creating %distance dependent 
%entropy in a field in contact with materials. 
In this Letter we study the entanglement of a quantum radiation field interacting with a dielectric medium. 
In particular, we describe the quantum mixed state of a field interacting with a dielectric through plasma and Drude models, and show that these generate very different entanglement behavior, as manifested in the entanglement entropy of the field. We also present a formula for a ``Casimir'' entanglement entropy, i.e. the distance dependence of the field entropy. Finally, we study a toy model of the interaction between two plates. In this model, the field entanglement entropy is divergent, however, as in the Casimir effect, its distance-dependent part is finite and the field matter entanglement is reduced when the objects are far.
\end{abstract}
\maketitle

%The nature of interaction of quantum radiation fields with matter is at the heart of many questions in physics. From a practical point of view, the effect of the matter on the long wavelength field modes is conveniently encoded by means of the dielectric response of the material \cite{LP}, allowing the calculation of various response functions. From a fundamental perspective, it is also interesting to understand the mixed nature of the quantum state of such a field. %

The theory describing electromagnetic fluctuations in a linear material medium is an essential tool in understanding matter-radiation interaction.
% a corner stone of many developments in physics. 
Classically, thermally excited fluctuations as well as purely quantum, zero temperature, fluctuations in macroscopic media have been widely studied in terms of susceptibilities (see e.g. \cite{LP, levin1967theory}).
%Usually, the field matter at long wavelength is conveniently encoded
%Usually, the field matter at long wavelength is conveniently encoded by means of the dielectric susceptibility of the material \cite{LP}, and it's relations with various response functions.
%A prominent example of such a situation, is the Casimir effect, where the parametric dependence of the ground state energy of a field interacting with material boundaries gives rise to a force.
%A convenient approach is to encode the particular properties of the material by means of the dielectric response of the material \cite{LP}. 
%Different dielectric response functions may affect the field in different ways, and indeed the Casimir effect is sensitive to the frequency response of the material. 
%In a real material, the dielectric function must have an imaginary component, related to dissipative processes in the system. Thus 
%In a real material, the dielectric function is frequency dependent, an unavoidable consequence of the radiation matter coupling. 
While the use of such response functions have been highly successful in describing various phenomena, the actual quantum state of the field is usually not described.
Indeed, even at zero temperature, a field interacting with realistic materials (as opposed to idealized Dirichlet or Neumann boundaries), is in general not in a pure state and should be described in terms of a mixed state density matrix.  In modern terms, this mixed state is viewed as a consequence of matter-field entanglement, and carries a nonvanishing von-Neuman entropy, sometimes called ``entanglement entropy'' (EE). 

In this Letter we describe the quantum state associated with a field in a medium {\it given a dielectric function $\epsilon$}. The effective action for such a field includes frequency dependence. Usually, such frequency dependent action is associated with response functions, however, here, we have to understand the action as specifying an {instantaneous} form for the photonic density matrix. We do this by describing the effective density matrix of the field. % and the entanglement entropy.
%We do this by studying the equal time correlation functions.
%Entanglement in many body systems has been studied extensively in recent years,  in condensed matter, quantum information and field theory contexts.  
%Moreover, even when the dissipation is very weak, but temporal dispersion is present, as in the case of transparent medium, we should expect the radiation to be entangled with the mater.

As a measure of the field-matter entanglement, we use the entanglement entropy (EE) of the field. 
EE has been the subject of intense  investigations in recent years,  in a large variety of quantum systems, with applications to quantum information, condensed matter theory and high energy theory  (For a review see e.g. \cite{RevModPhys.80.517}). In particular, 
the entropy of radiation coupled to matter has been considered in numerous works. Usually the focus is on a single degree of freedom coupled to a bath of oscillators: For example, the entropy of a spin in the spin-boson model within the frame work of the Caldeira-Legget model \cite{caldeira1981influence} was considered in \cite{hur2008entanglement} while the entanglement of a single radiation mode with an array of spins was studied in \cite{lambert2004entanglement} for the Dicke model . On a different note, the EE of spatially separated intervals of vacuum (or ground state of a spin chains) has been considered in \cite{marcovitch2009critical,calabrese2009entanglement}. Here we consider a situation distinct from these works, and more akin to the scenario considered in studies of Casimir and Van der Waals interactions between macroscopic bodies: that of a field in contact with macroscopic dispersive bodies.

Consider the electromagnetic field interacting with a dielectric medium at zero temperature. We assume that no external charges or currents are present and work in the gauge $A_{0}=0$.
%In the long wave length limit, the field effective action 
The long wave-length effective action, 
\begin{eqnarray}\label{actionA}
S={1\over 4\pi}\int \D^3x{\D\omega}{\bf A}_{\omega}^*({\bf x})[{\omega}^2\epsilon({\omega},{\bf x})-\nabla\times\nabla ]{\bf A}_{\omega}({\bf x}),\,
\end{eqnarray}
%in the presence of the material is depends on the permittivity
encodes the field interaction with the material through the dielectric response function $\epsilon({\omega},{\bf x})$. As we show below, the appearance of frequency dispersion in $\epsilon({\omega},{\bf x})$ means that an action like \eqref{actionA} does not describe the state of the field as a ground state of a field Hamiltonian; indeed, the action is {nonlocal} in time and so cannot be quantized to yield a proper quantum Hamiltonian.
%, described by the effective action:
%
%The essential material properties are assumed to be encoded in $\epsilon$. 

%Here, we explore the nature of this mixed state, 

% the entanglement between the field and a dielectric material, and quantify the part of the entanglement due to dissipation in the system in terms of the entanglement entropy of the field ${\cal S}_{field}=-\Tr\rho\log\rho$, where $\rho$ is the field density matrix. We consider a toy model of the electromagnetic field for such interaction. %We find a cutoff divergent, part of the entropy, as well as a distance dependent part, which is cutoff independent. 

%In this Letter we study the nature of the mixed state associated with the field {\it given a dielectric function $\epsilon$}. Usually, such frequency dependent action is associated with response functions, however, here, we have to understand the action as specifying an {\it instantaneous} form for the photonic density matrix. We do this by studying the equal time correlation functions.

Clearly, one possible effect of coupling to the environment may be thermalization of our field. It is therefore tempting to try and describe the instantaneous state of such a field as effectively thermal i.e. $\rho\sim Z^{{-1}}e^{-\beta H_{eff}}$ for a reasonable effective field Hamiltonian $ H_{eff}$.  Such effective ``Entanglement Hamiltonians'' have been the focus on recent studies due to their relation to the conformal edge spectra of fractional quantum hall systems \cite{li2008entanglement,qi2011general}. 

We show, however, that in the field-matter system, the field cannot be viewed as simply thermal. Rather, different ranges of momenta feel different effective temperatures. %: Large $k$ photons are weakly interacting with the medium and so are less likely to be thermalized. Somewhat surprisingly, soft photons also seem to experience low effective temperatures. %, however their occupation number diverges at low momenta.  %, and what is the role of the frequency dispersion in the entropy. 
While the occupation probability of UV photons vanishes with high momenta, we still find that the field in \eqref{actionA} carries a logarithmic UV divergent quantum ``zero point'' entropy and show it's cutoff dependence in \eqref{Results entropy cutoff}.

Finally, we consider, {\it a la} Casimir, the distance dependence of the entropy associated with field interaction with a pair of objects separated by vacuum, and show a toy model in which the ``Casimir'' entanglement entropy is UV finite and decays with distance. Physically, this means that the field becomes more entangled with the plates as they come closer.
 
 %is problem for infinite systems as well as the distance dependence of the entropy associated with field interaction with a pair of objects separated by vacuum. %, as in the Casimir/Lifshitz situation.

%The aim of this paper is to understand the role of the temporal dispersion of the dielectric function in stating the entropy of a field.

%The action of the electromagnetic field can be written as:... 

Here we study a simplified scalar field version of the electromagnetic field action \eqref{actionA}. %We study a simplified scalar model described by
%As a simplified model, consider the action: 
\begin{eqnarray}\label{action}
S={1\over 4\pi}\int \D^3x{\D\omega}\phi_{\omega}^*({\bf x})[{\omega}^2\epsilon({\omega},{\bf x})-\nabla^2]\phi_{\omega}({\bf x}).
\end{eqnarray}
%The action \eqref{action} serves as .
%in the presence of a dielectric medium , represented by the permittivity $\epsilon$.
% defined by the wave equation ({\bf better start with action}):
%\begin{eqnarray}
%\nabla^2\phi+\omega^2\epsilon({\bf x},\omega)\phi=0
%\end{eqnarray}
Note that the UV dependence study bellow is valid for the full electromagnetic problem, since in homogenous systems  \eqref{actionA}  separates naturally into two %scalar fields associated with the two 
scalar fields describing the two polarization modes.
When the permittivity is independent of $\omega$, the action is local in time, and one can easily quantize the associated scalar action
{assuming} the conjugate momentum $\pi_{\phi}$ can be expressed in terms of $\dot{\phi}$ and doesn't depend on external fields. 
Such an action follows from the %a (Hermitian) field
Hamiltonian $H={1\over 4\pi}\int\D^3x [{\pi^{2}\over \epsilon({\bf x})}+(\nabla\phi)^{2}]$. It describes, at zero temperature, a pure state, and as such will have no entropy \footnote{The $x$ dependence of $\epsilon$ does not spoil this property}. %: it just means that the field has spatially non-uniform mass.

%then this is obviously an equation of a free field, has no temporal response, and should be associated with a system with a Hamiltonian, so that, in particular, at $T=0$ the system should have no entropy. 

The situation is fundamentally different if $\epsilon$ is $\omega$ dependent. The non-locality of the action \eqref{action} in time signals that our system is coupled to external degrees of freedom which have been integrated out, yielding a non trivial temporal response kernel.  In such a case, the system cannot be in a pure state %even at zero temperature, 
implying that our radiative system is entangled with the matter fields.

To proceed, let us briefly review the method of calculating entropies of Gaussian states (For details see, e.g. \cite{botero2003modewise,plenio2007introduction}). The calculation is facilitated by the fact that all information lies in the two point functions of the field.  For a scalar field with $n$ degrees of freedom $\phi_{n}$ and conjugate momenta $\pi_{n}$, one defines the vector:
%\begin{eqnarray}
$(O_1,..O_{2n})=(\phi_1,\pi_1,..\phi_n,\pi_n)$.
%\end{eqnarray}
The state of the field is then determined by the covariance matrix:
\begin{eqnarray}\label{Covariance Mat}
\gamma_{jk}=2Re\la(O_{j}-\la O_{j}\ra)(O_{k}-\la O_{k}\ra)\ra.
\end{eqnarray}
 $\gamma$ can be brought into a Williamson normal form $W\gamma W^{T}=diag(\mu_{1},...\mu_{n},\mu_{1},...\mu_{n})$ by means of a symplectic transformation $W\in Sp(2n)$ preserving the canonical commutation relations $[O_j,O_k]=i\sigma_{j,k}$, where $\sigma$ is the $2n\times2n$ matrix $$\sigma=\oplus_{j=1}^n \left(
\begin{array}{cc}
0 &  1  \\
-1  &  0
\end{array}
\right)~~~~;~~~~ W^t \sigma W=\sigma.$$
 \begin{comment} %  (i.e. $W^t \sigma W=\sigma$ ).
The canonical commutation relations are $[O_j,O_k]=i\sigma_{j,k}$, where $\sigma$ is the $2n\times2n$ matrix $$\sigma=\oplus_{j=1}^n \left(
\begin{array}{cc}
0 &  1  \\
-1  &  0
\end{array}
\right).$$
 \end{comment} 
% preserving $\sigma$.
%To compute the entropy 

The occupation probability of the normal modes are directly related to the symplectic eigenvalues $\mu_{i}$ of $\gamma$. 
%($\mu_{i}$ are also equal to the positive eigenvalues $\mu_{i}$  of ${i\over\hbar} \sigma \gamma$ (check signs). ) 
Finally, the density matrix can be written as:
\bea\label{Effective rho}
\rho_{}=Z^{-1}e^{-\sum_{i} u_{i}\Phi^{+}_{i}\Phi_{i}}~;~u_{i}=\log{\mu_{i}+{1}\over \mu_{i}-{1}}
\eea
where $\Phi^{+}_{i}$ is a bosonic creation operators for the normal mode $i$. 
%Having found the $\mu_{i}s$, 
The entropy is given by:
\begin{eqnarray}&\label{Entropy formula}
{\cal S}=\sum_{i} h(\mu_{i}) ~;~ h(\mu)={\mu +1\over 2}\log{\mu +1\over 2}-{\mu-1\over 2}\log{\mu-1\over 2}
\end{eqnarray} 

%{\bf Remark:} Not every correlation matrix is consistent with Gaussian state. as explained it has to obay $\gamma-i\sigma\geq 0$ (check)

%{\bf Remark:}  For a discrete system we will have different relation of k and omega.

%While carrying the symplectic transformations needed to diagonalize \eqref{Covariance Mat} implied by our action \eqref{action}, is a hard task, the all important case of a homogenous system allows us to proceed analytically. 
An additional, technical simplification occurs, if we assume that no $\la\phi\pi\ra$ correlations are present. It is then known that the symplectic eigenvalues are square roots of eigenvalues of $\Gamma={\cal G}{\cal H}$, where ${\cal G}$ and ${\cal H}$ are field and field momentum two point functions, respectively. %, a property which we will utilize here. 
%Next, we study the properties of the field state defined by our action \eqref{action} in a homogenous system. 

For the translationally invariant medium the symplectic eigenvalues can be labeled by momentum and given by $\mu_{k}=2\pi^{-1}(g_{{\bf k}}h_{\bf k})^{{1/2}}$, where $g_{{\bf k}}h_{\bf k}$ is the product ${\cal G}{\cal H}$ in ${\bf k}$ space given by $g_{k}=\la \phi^2\ra_{\bf k}=\int \D {\bf x} e^{{i\bf x\cdot\bf k}}\la\phi({\bf x},t')\phi({\bf 0},t)\ra|_{t'\rightarrow t_{+}}$ and similarly  $h_{k}=\la \pi^2\ra_{\bf k}$.  
The field entropy {\it per unit volume} can be written as:
\begin{eqnarray}\label{S field}
{{\cal S}_{field}}=-{ V}^{-1}\Tr \rho_{field}\log\rho_{field}=\int \D{\bf k}h(\mu_{\bf k})
\end{eqnarray}
For concreteness, let us choose a typical dielectric function $\epsilon$ to use in \eqref{corrphi} and \eqref{corrpi}.
%Consider a dielectric function of the form 
We take: $\epsilon({\omega},{\bf x})=1+4\pi\chi(\omega,{\bf x})$. 
%We immediately see a difference between two popular effective permittivities used in calculations of Casimir forces. 
With a typical susceptibiliy of the form $\chi_{b}={{\omega_{p}^{2}}\over (\omega_0^{2}-\omega^2-i\gamma_p \omega)}$ (for a conductor we will add a Drude function $\chi_{c}= \chi_b+i{\omega_{c}^{2}\over \omega(\gamma_c-i \omega)}$). % (where the conductivity is given as usual by $\sigma={ne^2\over m}{1\over  (\gamma_c-i \omega)}$).% ( {\bf remark:} these definitions are in SI from jackson, and so lack the $4\pi$ factor from cgs used in my definitions here). 
 The action \eqref{action} allows us to compute, per $k$ mode:
\begin{eqnarray}\label{corrphi}
g_{k}=
\int_0^{\infty} \D\omega 
 {1\over  {\omega^2   \epsilon({\bf k},i\omega)+k^2}}
\end{eqnarray}
 and, using time point splitting, 
 \begin{eqnarray}\label{corrpi}
h_{k}= 
\int_0^{\infty} \D\omega 
 {1\over  {\omega^2    \epsilon({\bf k},i\omega)+k^2}}(k^2+4\pi\omega^2\chi(i|\omega|)).
\end{eqnarray}
%These correlations can now be used to compute the entropy as follows.
Using, the integrals \eqref{corrphi},\eqref{corrpi} and \eqref{Effective rho} we represent the density matrix as $\rho\sim exp(-\sum_{\bf k} u_{{\bf k}}\Phi^{+}_{{\bf k}}\Phi_{{\bf k}})$.

\begin{comment}
The density matrix of the field can be written as:
\bea
\rho_{field}=Z^{-1}e^{-\sum_{k} u_{k}\Phi^{+}_{k}\Phi_{k}}~;~u_{k}=\log{\sqrt{\frac{4}{\pi ^2}g_kh_k}+{1}\over \sqrt{\frac{4}{\pi ^2}g_kh_k}-{1}}
\eea
where $\Phi^{+}_{k}$ is a bosonic creation operator at momentum $k$. 
\end{comment}
One may try to interpret this state as thermal, i.e.: $\rho\sim exp(-\beta H_{eff})$.
The effective field Hamiltonian is then given by: $H_{eff}=\int\D{\bf x}\D{\bf x}'  \hat{u}({\bf x}-{\bf x}' )\Phi^{+}_{{\bf x}}\Phi_{{\bf x}'}$ where 
$\hat{u}=\beta^{-1} \int\D{\bf k} u_{\bf k}e^{i {\bf k}\cdot({\bf x}-{\bf x}' )}$.
However, we find that generically
the fourier transformed $\hat{u}$ gives us a non-local $H_{eff}$. %The decay properties of this $H_{eff}$ will be described elsewhere.

Alternatively, we may interpret $u_{\bf k}$ as $u_{\bf k}=\beta_{k}E_{k,free}$ where $E_{k,free}$ are the photon energies without the interaction. Thus, different momenta ${\bf k}$ feel different effective temperatures. %: for example, high momenta, for which the material is essentially transparent are mostly un-thermalized while ${\bf k}$s with strong dielectric response may feel a significant effective temperature. 
Estimating $u_{\bf k}$ from the integrals \eqref{corrphi},\eqref{corrpi}, we find for 
 the soft modes, $u_{\bf k}=\beta_{k}E_{k,free}\propto \sqrt{k}$ and so the effective temperature:% goes as:
$$
T_k\equiv{k\over \beta_{k}}\propto {\sqrt{k}}~~~as~~~~k\rightarrow 0.
$$
The energy and number of occupied soft modes per unit volume up to a given $k_m$ are proportional to $k_m^{d+1/2}$ and $k_m^{d-1/2}$, respectively, are finite and small. The number variance of modes up to  $k_m$ is $\la \delta N^2\ra\sim \int \D^d kn_{k}(1+n_{k})\sim k_m^{d-1}$ in $d>1$. However, for $d=1$ we find a curious infra-red divergence: $\la \delta N^2\ra=-\log \varepsilon_I$, where $\varepsilon_I$ is an infra-red cutoff, inversely proportional to the system size.

%Bellow 
Next, we find that the field entropy \eqref{S field} suffers from a UV divergence.
%To compute the entropy, we estimate the symplectic eigenvalues $\mu_{\bf k}$ %using eq \eqref{corrphi},\eqref{corrpi} 
To do so, we estimate the symplectic eigenvalues $\mu_{\bf k}$ 
to lowest order in $\omega_{p}, \omega_{c}$ and use \eqref{S field}. %, the full quantum von-Neumann entropy of the field per unit volume. 
 We find that $\mu_{\bf k}\sim 1+ \pi^{-1}{2\omega_{p}^{2}\gamma_{p}\log k }+..$ for $k\gg1$ , and integration over momenta yields %the entropy diverges at the UV, with a cutoff dependence  
 (in 3d, $\hbar=c=1$ units):
\bea\label{Results entropy cutoff}&
%{{\cal S}_{field}\over Vol}=-{1\over Vol}\Tr \rho_{field}\log\rho_{field}\sim \nn & 
{\cal S}_{field}=\Big\{
\begin{array}{cc}
(\omega_{p}^{2}\gamma_p+\omega_{c}^{2}\gamma_c)\log(\Lambda)^{3}  &     \\
\omega_{p}^{2}\omega_{0}\log(\Lambda)^{2}  &   \gamma_{c}=\gamma_{p}=0  \\
 0 &    plasma\,\, model
\end{array}, 
\eea
%where $\rho_{\phi}$ is the density matrix of the field $\phi$, 
where $\Lambda$ is a high momentum (UV) cutoff. Numerically, the approximations used in \eqref{Results entropy cutoff} actually recover the correct $\Lambda$ dependence even for large values of $\omega_{p}, \omega_{c}$, since the dielectric response decays at the large $k,\omega$ limit.

%This highlights the different roles of generating quantum entropy, where the dissipative terms, describing interaction with additional degrees of freedom, are dominant. However, even when they are absent we still have entanglement between the field and the dielectric.

It is interesting to observe the special place of the  {\it "pure plasma"} limit response function. %({\bf Real or imaginary frequencies?}). 
We can easily understand the result \eqref{Results entropy cutoff} as follows.  An idealized plasma simply adds a finite mass to the field inside the region it occupies, completely expelling frequencies smaller than $\omega_{p}$.  Explicitly, substituting the plasma permittivity limit form $\chi=- { \omega_{c}^2\over \omega^2}$ in the action \eqref{action}, %in a homogenous space, 
%we see that $\chi$ 
it produces a mass term for $\phi$. Thus, the resulting action is consistent with a Hermitian field Hamiltonian, and as such, at zero temperature, to a pure state. % \new The result is expected on physical grounds: %One can also understand the vanishing of entropy for pure plasma as follows:  A pure plasma essentially 
 %Consider a slab of material, with a pure plasma  dielectric response. Such 
%An idealized plasma %has a complete: The system will be completely transparent to radiation above the plasma frequency, and completely reflecting at lower frequencies. % The field will 
%adds a finite mass to the field inside the region occupied by the plasma. % and will not generate entropy. 
%\old
%({\bf say something about degeneracies?}). 
Interestingly, the use of a pure plasma in the computations of Casimir energy and Entropy has been at the heart of a recent debate \cite{bezerra2004violation,decca2005precise,brevik2005temperature,Sushkov:2011fk}.  %We do not make any claims regarding the debate, but 
note that the distinction between the Casimir entropy in the two models is manifested in the full quantum entropy computed herein \footnote{It is possible to incorporate the high momentum cutoff more naturally by including spatial dispersion in
$
\epsilon (i \omega
   ,k)=\epsilon
   _0\left(1+\frac{f}{
   \text{A k}^2 +\gamma  \omega +\omega
   ^2+\omega _0^2}\right).
$
This expression does not appear to significantly affect cutoff dependence.}.

Having shown that the entropy ${{\cal S}_{field}}$ is UV divergent, it is natural to ask, in analogy with the Casimir effect, what is the distance dependence of the entropy of interaction with two distinct bodies $A$ and $B$? Is it UV finite? %and is the distance dependent part of it is UV finite. 
To answer such questions, we define:% a ``Casimir EE''
%S_{{\phi}}(A,B)=dist.~depend.~part~of~-\Tr \rho_{\phi}\log\rho_{\phi}=
\bea{\cal S}_{{R}}(A,B)={\cal S}(A\cup B)-{\cal S}(A)-{\cal S}(B),\eea%+{\cal S}(\emptyset)\, ,
%for a medium described by $\epsilon=\epsilon_{0}+4\pi\chi_{A}+4\pi\chi_{B}$ where the responses $\chi_{A},\chi_{B}$ are non-vanishing only on the separate bodies $A$ and $B$, respectively.
%The entropy of the field may be quantified in various ways. Here we use as a measure the von-Neumann of the field. Thus for a pair of bodies $A$ and $B$ we define
%\begin{eqnarray}
%S_{{\phi}}(A,B)=dist.~depend.~part~of~-\Tr \rho_{\phi}\log\rho_{\phi}
%\end{eqnarray}
%where $\rho_{\phi}$ is the density matrix of the field $\phi$, in the presence of a body $A$ and $B$. 
%For the full electromagnetic problem this calculation is challenging. As we will see, even for a Gaussian, scalar field it is very hard to make progress analytically and we have to resort to numerics. 
${\cal S}_{{R}}$ should not be confused with the %more familiar 
Casimir entropy, defined as ${\cal S}_{C}(A,B)=-\lim_{T\rightarrow 0}\partial_{T}F_{C}$, where $F_{C}$ is the Casimir free energy, obtained by subtracting all distance independent terms from the free energy of the EM field in the presence of the bodies $A,B$. % (or boundary conditions). 
Also, ${\cal S}_{{R}}$ is distinct from the "relative entropy" of probability theory, as we are comparing different systems, and not merely different statistical information about the same system.

%It is important to note that, while the sub-additivity of von-Neumann entropy \cite{araki1970entropy} shows that ${\cal S}_{R}$ is larger than the total entropy of the field and matter, this is no longer evident once subtractions are taking place in order to define ${\cal S}_{R}$ and ${\cal S}_{C}$.

The relevance of Casimir entropy ${\cal S}_{C}$ to understanding thermal corrections of the Lifshitz formula has been pointed out in many papers (see e.g. \cite{bezerra2004violation,Geyer:2005ap,bordag2010casimir}), where it was noticed that as $T\rightarrow 0$, ${\cal S}_{C}$ may not go to zero when using the Drude model (as might be expected by the Nernts theorem). It is interesting to note that while the Casimir EE ${\cal S}_{{R}}$ is distinct from ${\cal S}_{C}$, a similar behavior is observed in \eqref{Results entropy cutoff}.
%${\cal S}_{C}$ has a clear thermodynamic meaning, especially at high temperatures, where the Casimir force is entirely entropic \cite{feinberg2001casimir}.
%Indeed, at 
In addition, at high temperatures we expect ${\cal S}_{R}={\cal S}_{C}$, as most of the field entropy will be thermal (Technically, the relevant Green's function gets its major contribution from the $\omega=0$ Matsubara pole), in addition, the Casimir force is entirely entropic \cite{feinberg2001casimir}.

 %Our aim is to understand if ${\cal S}_{R}(A,B)$ is indeed cutoff independent. We proceed by defining the needed operators. This problem is significantly more complicated than computing the Casimir energy  ({\bf after deciding move this sentance}).

%Without carrying out subtractions, the expression for $S_{\phi}$ is divergent, thus, similarly to the Casimir energy, we define the distance dependent part of the entropy of the field as:
%\begin{eqnarray}
%S_{{R}}(A\cup B)=S_{{\phi}}(A\cup B)-S_{{\phi}}(A)-S_{{\phi}}(B)+S_{{\phi}}(\emptyset)
%\end{eqnarray}

%\begin{comment}

In recent years, numerous results for Casimir interaction between objects have been obtained using a TGTG representation of the energy \cite{kenneth2006opposites}. In these methods one separates between $T$ operators associated with local properties of each body, and a free green's functions interpolating between them. See, e.g. (\cite{emig2007casimir,milton2008multiple,kenneth2008casimir}). 
A natural question is: Can we describe ${\cal S}_{R}$ in similar terms? %Is there a ``TGTG'' type formula for it? % (\cite{kenneth2006opposites}) ? 
Using the methods of \cite{kenneth2008casimir}, assuming that $\phi\pi$ correlators vanish, we find an analogous expression for  ${\cal S}_{R}(A,B)$: 
\bea \label{S relative}&\label{SrelativeTGTG}
{\cal S}_{R}(A,B)=-\frac{1}{\pi }\int _{1/2-\text{i$\infty $}}^{1/2+\text{i$\infty $}}dx\frac{\log\left[\frac{\sqrt{x}+\frac{1}{2}}{\sqrt{x}-\frac{1}{2}}\right]}{2 \sqrt{x}}\Tr\log\nn &\left(1-\frac{1}{1-K_A}\left(K_AK_B+K_{\text{AUB}}-K_A-K_B\right)\frac{1}{1-K_B}\right)\eea
%. Indeed, 
%It is possible to derive a general expression for ${\cal S}_{R}(A,B)$ \old \footnote{In preparation}: %along the lines used the derivation of the  $TGTG$ expression for Casimir energy \cite{kenneth2008casimir}. 
Here $
K_A= \frac{1}{x-1/4+i s}\left(\Gamma_{A}-1/4\right)
$ is  defined via $\Gamma(x,x')=i\int\la\phi(x)\phi(x'')\ra\la\pi(x'')\pi(x')\ra\D x''$ %, where $\pi$ is the conjugate momentum operator to $\phi$, 
%is the operator product of field and momentum covariances.  %\new shouldnt we have 1/4 in the integral limits? probably fine: we use here factor 2 in the definition of the correlation matrix.\old
, and similar expressions hold for $K_{B},K_{AUB}$.
%\new 
Expression \eqref{S relative} is similar to the TGTG formulas in it's form: an integral over the TrLog of a combination of Green's functions. It differs from such formulas in several aspects: The integration variable $x$ is not a frequency variable, but rather an auxiliary spectral variable, the presence of the term $K_{AUB}$ doesn't allow for full separation into local object properties and free propagators and the non-analyticity at $x=1/4$. All of these make the formula harder to use than the TGTG formulas. Nevertheless, it is a useful starting point for multiple scattering expansions.%when the bodies are weakly entangled with the field, so that $||\Gamma_{A}-1/4||\ll 1$.\old %$\old
%Despite the elegance of \eqref{S relative}, it is hard to work with because of the non analytic nature of the integral argument close to $x=1/4$. %Unfortunately, this non analyticity often plagues entanglement entropy calculations. 

%\end{comment}
Here, we defer a detailed study of the formula \eqref{S relative} and proceed instead to study the dependence of entanglement on the distance between bodies in a simple toy model:
%It is not clear if there is an infrared problem with this expression. 

%To lowest order, if the trace of $\Gamma-1/4$ exits, we can write: $S\sim f(\epsilon)\Tr(\Gamma-1/4)$ for an appropriate f, where $\epsilon$ some kind of IR cutoff %\old 

\begin{figure}
\includegraphics[scale=0.25]{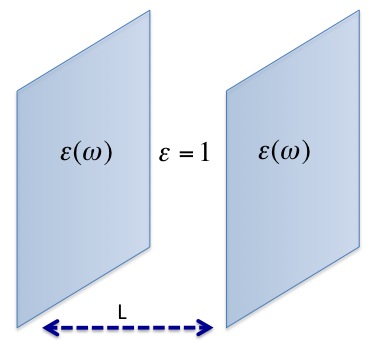}\caption{``Casimir'' entanglement entropy ${\cal S_{R}}$ of plates.}\label{Planes}
\end{figure}
%To study the dependence of entanglement on the distance between the bodies, we 
Consider a field interacting with a medium, defined through the following (Wick rotated) action:
\bea\nonumber&\label{Action interaction}
S_{\text{pure}}=\frac{1}{2}\int _{-\infty }^{\infty }\frac{d\omega }{2\pi }\Big\{\int d^3x[\phi ^*\left(-\omega ^2+\nabla ^2\right)\phi _{\omega }]-\nn &\int_{B} d^3x[\frac{\left(\omega ^2+\omega _0{}^2\right)}{8\pi ^2}\psi _{\omega }{}^*\psi _{\omega }+\omega _p\omega ( \psi _{\omega }{}^*\phi _{\omega }+  \psi _{\omega }\phi _{\omega }{}^*)]\Big\}\eea
where $\psi$ is a matter field which is confined to the body(s) $B$. This action may be viewed as a {\it purification} of the state $\rho_{\phi}$ of $\phi$,  i.e. we write our state as $\rho_{\phi}=\Tr_{\psi}|\Omega\ra\la\Omega|$ for some pure state $|\Omega\ra$ in the larger $\phi,\psi$ space.
The action \eqref{Action interaction} corresponds to the form $\chi (\omega )=\frac{\omega _p{}^2}{\omega _0{}^2-\omega ^2}$ of the response of $\phi$ to a transparent, but dispersive medium.

Finally, to render the problem exactly solvable, we will assume that the $\phi\psi$ coupling in \eqref{Action interaction} comes from the form $\omega_p\int\psi\dot{\phi}$ in the classical Lagrangian. Note, that while on the classical level this Lagrangian coupling is equivalent to $-\omega_p\int\phi\dot{\psi}$, the actual quantum mechanical transformation required to affect this for gauge fields is of the Power-Zineau-Wooley (PZW) type used to interchange the minimal coupling ${\bf A\cdot J}$ to ${\bf E\cdot P}$ in quantum electrodynamics \cite{cohen2004photons}. However, PZW mixes radiation and dipole degrees of freedom and does not preserve entanglement properties.

%The situation with non homogenous dielectric is much more complicated. Here, we consider field interacting with a dielectric body (or bodies) $B$. 

%Here we will study the simplified form $\chi (\omega )=\frac{\omega _p{}^2}{\omega _0{}^2-\omega ^2}$. With this form we choose a {\it purification} of our state, i.e. write our state as $\rho_{\phi}=\Tr_{\psi}|\Omega\ra\la\Omega|$ for some pure state $|\Omega\ra$ in the combined system of the field $\phi$ and auxiliary field $\psi$. After a Wick rotation,
%the action in the combined system is:
%\bea\nonumber&
%S_{\text{pure}}=\frac{1}{2}\int _{-\infty }^{\infty }\frac{d\omega }{2\pi }\Big\{\int d^3x[\phi ^*\left(-\omega ^2+\nabla ^2\right)\phi _{\omega }]-\nn &\int_{B} d^3x[\frac{\left(\omega ^2+\omega _0{}^2\right)}{8\pi ^2}\psi _{\omega }{}^*\psi _{\omega }+\omega _p\omega ( \psi _{\omega }{}^*\phi _{\omega }+  \psi _{\omega }\phi _{\omega }{}^*)]\Big\}\eea

\begin{comment}
\bea\nonumber&
S_{\text{eff}}=\frac{1}{2}\int _{-\infty }^{\infty }\frac{d\omega }{2\pi }\Big\{\int d^3x\left[\phi ^*\left(\omega ^2+\nabla ^2\right)\phi _{\omega }\right]+\nn & \int _Bd^3x\left[\frac{-\psi _{\omega }{}^*\psi _{\omega }}{8\pi ^2\omega _p{}^2 }\left(\omega ^2+\omega _0{}^2\right)+\omega  \psi _{\omega }{}^*\phi _{\omega }+\omega  \psi _{\omega }\phi _{\omega }{}^*\right]\Big\} \nonumber\eea
\end{comment}

To compute the entropy of the field $\phi$ more efficiently, we use that ${\cal S}_{\phi}={\cal S}_{\psi}$ (noting that \eqref{Action interaction} describes a pure state). ${\cal S}_{\psi}$ is obtained from the effective action for $\psi$:
\bea & S_{\text{eff}}[\psi ]=-\frac{1}{4\pi }\int_{B} d^3x d^3x'[\omega_{p}^{2}\omega^{2}\langle {\bf x}|G_{0}^{B}|{\bf x}'\rangle +\nn & \delta ({\bf x}-{\bf x}')\frac{\omega ^2+\omega _0{}^2}{8\pi ^2 }]\psi _{\omega }{}^*({\bf x})\psi _{\omega }{}^*({\bf x}'),  \eea
where $G_{0}^{B}$  is the restriction of the free Green's function $G_{0}=\frac{1}{-\omega ^2+\nabla ^2}$ to the bodies $B$.

%Denoting $G_{0}^{B}$ the Green's function $G_{0}$ restricted to the bodies, and for convenience discretizing space, we have the following formula for the two point function of $\psi$:% (computed in $B$):
The two point function of $\psi$ is:
\bea \label{X corr} \la\psi({\bf x}) \psi({\bf x}') \ra=\int_0^{\infty } \langle {\bf x}| \frac{1}{\frac{\omega ^2+\omega _0{}^2}{8\pi ^2}+\omega _p{}^2 \omega^2 G_{0}^{B}}  |{\bf x}'\rangle  { d\omega\over 2\pi}\eea
with a similar expression holding for the conjugate momentum of $\psi$ (obtained from the Lagrangian in $S_{\text{pure}}$):
\bea & \label{P corr} {\cal H}_{{\bf x}{\bf x}'}=\la\pi_{\psi({\bf x})}\pi_{\psi({\bf x}')}\ra=-{1\over 16\pi^{4}}\nn & \int_0^{\infty } \langle {\bf x}| \frac{\omega^{2}}{\frac{\omega ^2+\omega _0{}^2}{8\pi ^2 }+\omega _p{}^2\omega^2 G_{0}^{B}}-8\pi ^2 \delta({\bf x}-{\bf x}')  |{\bf x}'\rangle   { d\omega\over 2\pi}\eea

%To make analytical progress, we choose the bodies to be 
Next, we study the entropy generated by parallel planes separated by a distance $L$, as illustrated in Fig.\ref{Planes}. 
To control UV behavior we place the system on a lattice, and use the discrete analogue of the free Green's function (in 1D) as: $G_0(n,m,k_{\perp})=\frac{e^{-q|n-m|}}{2\sinh q}$, where: $\cosh q=1+\frac{\omega ^2+{k_{\perp}^{2}}}{2}$ ($k_{\perp}$ is parallel to the plates). %Additional dimensions are introduced by adding a ($k_{\perp}$ dependent) mass term to $\omega^{2}$ in the 1D Green's function.

We first consider the 1D case. In 1D, the planes reduce to a pair of sites, and the kernel  $(\frac{\omega ^2+\omega _0{}^2}{8\pi ^2 }+\omega _p{}^2\omega^2 G_{0}^{B})^{-1}$ in eqs. (\ref{X corr},\ref{P corr}) 
is a 2X2 matrix, amenable to an analytic treatment. %Under the conditions $\omega_{p}\ll 1$, $\omega_{0}>\omega_{p}$ and $L\gg 1$. 
Estimating the two point functions  \eqref{X corr} and \eqref{P corr} using Watson's lemma, we find, at large distances $L\gg 1$ (assuming $\omega_{0}>\omega_{p}$ and up to order $\omega_{p}^{2}$):% (showing only terms up to order $\omega_{p}^{2}$):
\bea \label{Gnm 2plates} & {\cal G}_{nm}=\la\psi(n) \psi(m) \ra\sim {1\over 2\pi}\left(
\begin{array}{cc}
 A_G & -\frac{32\text{  }\pi ^4 \omega _p{}^2 }{\omega _0{}^4L^2}  \\
 -\frac{32\text{  }\pi ^4 \omega _p{}^2 }{\omega _0{}^4L^2} \text{  } & A_G \end{array} \nonumber
\right); \nn &
A_G=\frac{4 \pi^3 }{\omega_0}-\frac{32 \pi ^4 \omega_p^2(\omega_0^2 \arccos[\frac{2}{\omega_0}]-2 \sqrt{\omega_0^2-4})}{\omega_0^2(\omega_0^2-4)^{3/2}}.
\eea

Similarly,  the momentum correlation function $ {\cal H}_{nm}$ is 
\bea \label{Hnm 2plates}  & {\cal H}_{nm}\sim  \frac{-1}{32\pi ^5}\left(
\begin{array}{cc}
 A_H & -\frac{192\pi^4 \omega_p^2 }{\omega_0^4 L^4}\\
 -\frac{192\pi^4 \omega_p^2 }{\omega_0^4 L^4} & A_H
\end{array}
\right); \\ \nonumber & A_H={\omega _0\over 4 \pi } +\frac{32 \pi ^4 \omega_p^2 \left(2 \sqrt{\omega_0^2-4}+(\omega_0^2-8) \arccos[\frac{2}{\omega _0}]\right)}{(\omega _0^2-4)^{3/2}}.\eea
The dominant terms in the product $\cal{GH}$ are proportional to $\theta\equiv\frac{1}{64 \pi ^6}A_{G}A_{H}$, and are independent of distance.

As expected on physical grounds, the entropy strongly depends on the pinning frequency $\omega_{0}$.
We note that the entropy is maximized when $\omega_{0}$ is small as possible, i.e. $\omega_{0}\sim \omega_{p}$. In this limit the $\psi$ field is only weakly restrained to $\psi=0$ values, and thus generates large entropy for the $\phi$ field.  
In the weak coupling limit $\omega _p\ll 1$ we find the eigenvalues of $\cal{GH}$ to behave as $\theta\left(\omega _0=\omega _p\right)\sim \frac{1}{4}+2 \pi \omega _p \log \frac{4}{e \omega _p}  $.
%$4^{-1}+2 \pi \omega _p \log{4}{e^{-1} {\omega_p}^{-1}}  $ .
In the opposite limit, when
$\omega _0\gg \omega_{p}$
the $\psi$ field is essentially ``pinned'' down, and we have   
$
\theta\sim \frac{1}{4}+\frac{8 \pi  \omega _p^2}{\omega _0^3}-\frac{8 \left(\pi ^2 \omega _p^2\right)}{\omega _0^4}.$ As $\omega_{0}\rightarrow \infty$ we recover the minimal possible value of the product ${\cal G}{\cal H}$ allowed by the uncertainty principle.

Finally, the symplectic eigenvalues of the covariance matrix are found, to leading orders in ${1\over L},\omega_{p}$, to be 
\bea \label{symp spect} \lambda _{\pm }=\sqrt{\theta}\pm A_{H}\omega _p^2  ({4 \pi^{2}\omega _0^4 \sqrt{\theta}  L^2})^{-1}\eea 
% \frac{1}{ \sqrt{\theta}}\frac{  A_{H}\omega _p^2 }{4 \pi^{2}\omega _0^4 L^2}
and the entropy %of the field $\phi$ (equaling that of $\psi$) is evaluated, 
using \eqref{Entropy formula} for large $L$ to %, by ${\cal S}_{\phi}=h(\lambda_{+})+h(\lambda_{-})$. 
%Expanding the entropy we find it 
behave as: \bea  {\cal S}_{\phi}= {\cal S}(L\rightarrow\infty)+32 \pi ^7 \omega _p^2 \omega_0^{-3} L^{-4}.\eea

At dimensions $D>1$, the momentum $k_{\perp}$ is a good quantum number. We proceed to compute the entropy per $k_{\perp}$ by estimating  $\la\psi(x,k_{\perp}) \psi(x',k_{\perp}) \ra$ and $\la\pi_{\psi(x,k_{\perp})}\pi_{\psi(x',k_{\perp})}\ra$. 
\begin{comment}
 we have to include integration over the transverse momenta $k_{\perp}$ with the 1D results. $k_{\perp}$ simply appears as an effective mass in $G_{0}^{B}$ when computing $\la\psi(x,k_{\perp}) \psi(x',k_{\perp}) \ra$ and $\la\pi_{\psi(x,k_{\perp})}\pi_{\psi(x',k_{\perp})}\ra$ in eq. (\ref{X corr},\ref{P corr}).\end{comment} %We find that the large $k_{\perp}$ behavior of the symplectic spectrum  \eqref{symp spect} is modified by $\theta \sim {1\over 4}+\frac{2 \pi {\omega_p}^2}{k_{\perp}{\omega_0}}$, and that integrating over $k_{\perp}$  yields  a UV divergent entropy for the system per unit area. However, the 
We find that the $L$ dependent, off-diagonal elements in (\ref{Gnm 2plates},\ref{Hnm 2plates}) are exponentially decaying  as $e^{-L |k_{\perp}|}$. Integrating over $k_{\perp}$ yields \bea {\cal S}_{\phi}={\cal S}(L\rightarrow\infty)+ {C_{d}(\omega_{0},\omega_{p})}{L^{-3-d}}\eea where  $C_{d}(\omega_{0},\omega_{p})$ is a constant.
% We find that $\partial_{L} {\cal S}_{\phi}$ exists even for the cutoff on $k_{\perp}$ going to infinity. The $k_{\perp}$ integration modifies the $L$ dependence to ${\cal S}(L\rightarrow\infty)+\frac{C_{d}}{L^{3+d}}$ where %asymptotics suggests $\alpha=d$ and 
% $C_{d}(\omega_{0},\omega_{p})$ is a constant.
%\subsection*{Summary}

{\it {Discussion:}} We studied the quantum state of a field in the presence of a dielectric material. 
We found that such a field is described by a density matrix whose Von-Neumann
entropy diverges as described by eq. \eqref{Results entropy cutoff}. The state cannot be considered as thermal, but rather the photons have a $k$ dependent effective temperature. 
%, which is zero at high and low momenta. 
The occupation number of modes behaves as $n_{k}\sim 1/\sqrt{k}$ per unit volume for "soft" photons. A similar situation may arise when considering the phonons in solid. 
This effect is reminiscent of the infra red problem in quantum electrodynamics where,  infinite numbers of soft photons are generated in transition amplitudes (see, e.g. \cite{kibble1968coherent}). The situation here is different in that the system is in equilibrium, however we are integrating over dipole transitions in the material.

To find long distance features, unaffected by UV divergence, we considered $\cal{S}_{R}$, the distance dependent part of the entropy of two objects. 
In eq. \eqref{S relative} we present a novel general formula for the "Casimir Entanglement Entropy". 
Finally, we considered a toy model for the entropy generated in the field due to interaction between two thin planes. % , by purifying the state and computing the entropy of the auxiliary field. 
In this model $\cal{S}_{R}$ shows a $L^{-(3+d)}$ decay:
Thus, the field matter EE is reduced when the objects are far. Further investigation is needed to determine %whether 
the 
power law  in more realistic models.% are retained for 
%natural $\int\dot{\phi}\psi$ coupling. and UV finite behavior

%We expect that much more insight into the mixed state of the electromagnetic field  may be gained using numerical means to study 
%how other factors, such as geometries and vector properties, affect the Casimir entanglement entropy and entanglement spectrum. 
The measurement of the full quantum entropy is, in general, quite hard, but possible to achieve in certain situations \cite{klich2006measuring, klich2009quantum, song2011entanglement, cardy2011measuring}. 
In the case of Gaussian fields with translational invariance, it can be computed from mode occupations. We mention that many of the properties discussed are also valid for other fields, such as phonons in a solid. For example, occupation numbers of phonons in ultracold atoms or trapped ion systems may be measurable in experiments \cite{dutta2011non, brahms2011optically}, and may reveal some of the features described here. 

Financial support from NSF CAREER award No. DMR-0956053 is gratefully acknowledged.

\bibliography{/Users/iklich/dropbox/Work/KlichBib.bib}
\end{document}